\documentclass[]{JFM-FLM_Au}

\usepackage{graphicx}
\usepackage{dcolumn}
\usepackage{bm}

\usepackage{mathrsfs}
\usepackage{amsmath,amsfonts,amssymb,wasysym}
\usepackage{color}

\DeclareMathOperator{\mH}{\mathcal{H}}

\DeclareMathOperator{\cT}{\mathcal{T}}
\DeclareMathOperator{\cL}{\mathcal{L}}
\DeclareMathOperator{\cU}{\mathcal{U}}
\DeclareMathOperator{\cD}{\mathcal{D}}

\def\di{{\rm d}}

\newcommand{\Bn}{B}
\newcommand\Bi{B}
\newcommand{\Aend}{A_{\rm{sat}}}
\newcommand{\Wi}{W}
\newcommand{\ucd}{\stackrel{\kern0.1em\triangledown}}
\newcommand{\bs}{\boldsymbol}
\newcommand{\ys}{y_{\mathrm{y}}}

\newcommand{\be}{\begin{equation}}
\newcommand{\ee}{\end{equation}}

\newcommand{\pd}[2]{\frac{\partial #1}{\partial #2}}
\newcommand{\od}[2]{\frac{{\rm d} #1}{{\rm d} #2}}
\newcommand{\odd}[2]{\frac{{\rm d}^2 #1}{{\rm d} #2^2}}
\newcommand{\pdd}[2]{\frac{\partial^2 #1}{\partial #2^2}}
\def\ctxx{{\check\tau_{\rm xx}}}
\def\ctxy{{\check\tau_{\rm xy}}}
\def\ctyy{{\check\tau_{\rm yy}}}

\def\txx{{\tau_{\rm xx}}}
\def\txy{{\tau_{\rm xy}}}

\def\Txx{{T_{\rm xx}}}
\def\Txy{{T_{\rm xy}}}
\def\Tyy{{T_{\rm yy}}}
\def\tyy{{\tau_{\rm yy}}}

\def\Yb{{\bar{Y}}}
\def\cTb{{\bar{\cal T}}}
\def\Yc{{\check{Y}}}

\def\half{{\mbox{$1\over2$}}}

\lefttitle{J.D. Shemilt, N.J. Balmforth and D.R. Hewitt}
\righttitle{Journal of Fluid Mechanics}

\title{Nonlinear evolution of instability in inertialess elasto-viscoplastic Poiseuille flow}

\author{James D. Shemilt\aff{1}, Neil J. Balmforth\aff{1} \and Duncan R. Hewitt\aff{2}}

\affiliation{\aff{1} Department of Mathematics, University of British Columbia, 1984 Mathematics Road, Vancouver, BC, V6T 1Z2, Canada
\aff{2}DAMTP, Centre for Mathematical Sciences, Wilberforce Road, Cambridge CB3 0WA, UK}

\corresau{James D. Shemilt, \email{shemilt@math.ubc.ca}}

\begin{document}
\maketitle

\begin{abstract}
  A widely used constitutive
  law for elasto-viscoplastic fluids (Saramito's model) predicts linear
  instability in inertialess pressure-driven channel flow.
  The instability arises due to the yield stress of the fluid
  and is strongest at the shortest streamwise wavenumbers, calling
  into question the physical validity of the constitutive model.
  Here, we show that the short wavelengths can be controlled
  by the addition of polymer stress diffusion and conduct
  two-dimensional numerical simulations to explore
  the nonlinear dynamics. On reaching finite amplitude, the
  instability is shown to generate spatio-temporally complicated
  states. 
  Fluctuations about the final mean state are pronounced near
  and between the yield surfaces that
  border an unyielded plug spanning the centre of the channel.
  The instability and transition arise for Weissenberg numbers
  of order unity and higher. 
\end{abstract}



\section{Introduction}
\label{sec:intro}

Yield-stress fluids flow only when subjected to a sufficiently large stress,
deforming like solids below that critical threshold. Such materials are
commonplace in geophysical, biological and industrial settings
\citep{annrev,coussot25}. In addition,
many yield-stress fluids deform viscoelastically
both below and above the yield stress, and much recent effort
has been aimed at developping constitutive laws to capture
`elasto-viscoplastic' material behaviour 
\citep{fraggedakis16,coussot25}.
The most widely used of these laws are Saramito's models
\citep{saramito07,saramito09}, which
switch at the yield stress
between a Kelvin-Voigt viscoelastic solid and an Oldroyd-B-type
viscoelastic fluid with nonlinear viscous behaviour.
The predictions of the
Saramito models for simple Couette flows have been compared
with rheometric measurements, with a mix of agreement
and some setbacks \citep{fraggedakis16,KDR}. 
Armed with Saramito's laws, modelling
studies have sought to characterize elastic effects in a number
of settings, including 3D printing, the mucus lining of airways,
bubble dynamics, particle transport and sedimentation
and thinning threads \citep{franca2024,franca26,erken_2022_elastoviscoplastic,
sanjay21,balasubramanian24,mosch,kordalis26,fraggedakis16,chaparian20,
zakeri25,vpthread}. 

Nevertheless, many studies explore the repercussions of elasto-viscoplastic
rheology in either highly simplified configurations (rheometric flows)
or in large-scale simulations of
relatively complex flows. The middle ground, of idealized configurations
that can be interrogated cleanly and precisely,
has been left relatively untouched. This is 
concerning since constitutive laws can introduce new features
into the flow dynamics that may not represent real material
behaviour, and motivates the
current study, in which we re-examine a relatively simple problem:
the pressure-driven channel flow of an 
elasto-viscoplastic fluids described by Saramito's model.

For this flow configuration, recent work \citep{evpoiy} has pointed out
notable consequences of adopting the Bingham version of
Saramito's model \citep{saramito07}. In particular,
uni-directional base flows can develop
discontinuities in the normal stresses and shear rate
as they evolve from their initial state. Worse, \cite{evpoiy}
detected a novel linear instability at zero Reynolds number, with
the awkward feature that the shortest
streamwise wavelengths had the highest growth rates. Although
similar results have been found for other shear-thinning
visco-elastic flows \citep{wilson99,bodiguel,poole16},
the instablity in elasto-viscoplastic Poiseuille flow
was crucially found to be associated with the regions in the base flow
that lay either at or below the yield stress (the plastically deforming
or plugged regions).
Although the growth rates reached a plateau
at high wavenumber, they did not diverge,
indicating that the model equations were not mathematically
ill-posed. However,
the implied sensitivity to initial condition and the potential
development of the shortest spatial scales is concerning and
leads one to question the physical validity of the constitutive model.

Another awkward feature of the analysis
conducted by \citet{evpoiy} was the need to deal with 
stress discontinuities in the uni-directional base state,
which present a significant obstacle to
linear stability theory. This led \citet{evpoiy} to retreat to a
high-Weissenberg-number limit to simplify the problem, which obscures
the generality of the results.
In the present study, we instead add polymer stress diffusion to the constitutive law, and detail how this addition 
deals with the stress discontinuities
and controls unstable short wavelengths.
Stress diffusion is commonly added
to ease many viscoelastic flow simulations
\citep[e.g., ][]{dubief23,lellep24,lewy26}, and has been argued
to be present physically on the grounds of kinetic theory. Stress diffusion has also been used in some recent elasto-viscoplastic flow simulations \citep{dzanic24,from25,abdullah26}.
Here, including diffusion also allows us to conduct
fully nonlinear, two-dimensional numerical simulations
covering a range of Weissenberg numbers, to assess
the elasto-viscoplastic instability more completely and explore
its consequences on the nonlinear flow dynamics.

\def\hp{{\hat{p}}}
\def\hu{{\hat{u}}}
\def\hv{{\hat{v}}}
\def\htau{{\hat{\tau}}}
\def\hkappa{{\hat{\kappa}}}
\def\hx{{\hat{x}}}
\def\hy{{\hat{y}}}
\def\th{{\hat{t}}}

\section{Model and methods}

\subsection{Mathematical formulation}

We consider the flow of an incompressible, elasto-viscoplastic
fluid down a two-dimensional channel of width $\mH$, driven
by a pressure gradient $-\Gamma$.
Ignoring inertia and describing
the channel geometry by the Cartesian coordinates $(\hx,\hy)$,
mass and momentum balance demand
\begin{equation}
  \nabla\cdot\bs{\hu}=0,\quad \nabla \hp = \mu_s\nabla^2\bs{\hu}
  + \nabla\cdot\bs{\htau}+\Gamma\boldsymbol{\hx},\label{nseqd}
\end{equation}
for the velocity $\bs{\hu}=(\hu,\hv)$, and polymer stress, $\bs{\htau}$,
where $\mu_s$ denotes the viscosity associated with a Newtonian
contribution to the total stress (a
``solvent viscosity'')
and $\boldsymbol{\hx}$ is the unit vector in the $\hx$-direction.
Here, the departure of the pressure from the
background, $-\Gamma \hx$, is denoted by $\hp$.
We write the Bingham version of
Saramito's constitutive law \citep{saramito07} as
\begin{equation}
  \lambda \left[ \frac{D \bs{\tau}}{D\th}
    - (\bs{\nabla}\bs{u})^T \cdot \bs{\htau}
    - \bs{\htau} \cdot (\bs{\nabla}\bs{u}) \right]
  + \max\left(1-\frac{\tau_Y}{\htau},0\right)\bs{\htau} =
  \mu_p (\bs{\nabla}\bs{\hu} + (\bs{\nabla}\bs{\hu})^T)
  + \hkappa\nabla^2\bs{\htau},\label{coneq}
\end{equation}
where $\th$ is time, $D/D\th = \partial/\partial\th + {\bs{u}\cdot\nabla}$
is the material derivative, $\lambda$ is the fluid relaxation time,
$\htau=\sqrt{(\htau_{xx}-\htau_{yy})^2+4\htau_{xy}^2}$
is the polymer stress invariant,
$\mu_p$ is the polymeric viscosity, $\tau_Y$ is the yield stress
and $\hkappa$ is the polymer stress
diffusion coefficient.
At the channel walls, $\hy=\pm\frac12\mH$, we impose no slip 
and no-flux conditions on $\bs{\tau}$,
and assume periodic boundary conditions in $\hx$
in a domain of length $\cL$.

\subsection{Dimensionless model}

We express the governing equations in dimensionless form
by introducing the scaled variables,
\be
t = \frac{\cU\th}{\mH}, \quad
(x,y) = \frac{(\hx,\hy)}{\mH}, \quad
(u,v) = \frac{(\hu,\hv)}{\cU}, \quad
(p,\bs{\tau}) = \frac{2(\hp,\bs{\htau})}{\Gamma\mH},
\ee
where
\be
\cU=\frac{\Gamma\mH^2}{2\mu}
\qquad {\rm and} \qquad
\mu = \mu_p+\mu_s.
\ee
Then, in component form, we arrive at the dimensionless model equations,
\begin{align}
\pd{u}{x} + \pd{v}{y} &=0, \label{conteq}\\
\beta\left(\pdd{u}{y} + \pdd{u}{x}\right)
+ \pd{\txx}{x} + \pd{\txy}{y}  - \pd{p}{x} &= -2 ,\\
\beta \left(\pdd{v}{y}+ \pdd{v}{x}\right)
+  \pd{\txy}{x}+\pd{\tyy}{y} - \pd{p}{y} &= 0 , \label{1.9}\\
  \Wi \left( \frac{D\txx}{Dt} 
  - 2\txx\pd{u}{x} - 2\txy\pd{u}{y}\right)
  + Y \txx
  &= 2(1-\beta)\pd{u}{x}
  + \kappa \pdd{\txx}{x} + \kappa \pdd{\txx}{y},   \label{1.11}
  \\
  \Wi
  \left(
  \frac{D\txy}{Dt} 
- \txx\pd{v}{x} - \tyy\pd{u}{y} \right)
+ Y \txy
&= (1-\beta)\left(\pd{u}{y}+ \pd{v}{x}\right)
+ \kappa \pdd{\txy}{x} + \kappa \pdd{\txy}{y},
\\
\Wi \left(
  \frac{D\tyy}{Dt} 
- 2\txy\pd{v}{x} - 2\tyy\pd{v}{y}\right)
+ Y \tyy
&= 2(1-\beta)\pd{v}{y}
+ \kappa \pdd{\tyy}{x} + \kappa \pdd{\tyy}{y},
  \label{1.13}
\end{align}
where
\be
\frac{D}{Dt} = \pd{}{t} + u \pd{}{x} + v \pd{}{y} 
,
\qquad
Y = \max\left(1-\frac{\Bi}{\cT},0\right) \label{Ydef}
\ee
and
\be
\cT = \sqrt{(\txx-\tyy)^2+4 \txy^2}.
\ee
Here,
\be
\Wi = \frac{\lambda \cU}{\mH}, \quad
\beta = \frac{\mu_s}{\mu}, \quad
\Bi = \frac{4\tau_Y}{\Gamma\mH} \quad {\rm and} \quad
\kappa = \frac{\hkappa}{\cU\mH}
\ee
denote a Weissenberg number, 
viscosity ratio, Bingham number and dimensionless stress
diffusivity, respectively,
and constitute four of the main dimensionless parameters
of the problem.
The boundary conditions on the walls are
\be
u=v=\pd{\txx}{y}=\pd{\txy}{y}=\pd{\tyy}{y}=0
\quad {\rm at} \quad y=\pm\frac12,\label{bcs}
\ee
with periodicity demanded over the domain $0\leq x \leq L$,
where $L=\cL/\mH$ is the dimensionless domain length.


Note that a different formulation of the equations is presented
by \citet{evpoiy}: they scaled the variables differently
in order to prepare the dimensionless model for an analysis
in the limit $\Wi\gg1$. The alternative formulation scales $x$
and $y$ in different ways, as well as using a different characteristic
timescale. Here, we scale $x$ and $y$ in the same manner,
leading to a more transparent formulation, and avoid taking
any large Weissenburg number limit.

\section{Uni-directional base flows and their linear stability}

\subsection{Start-up}

Uni-directional base flows, with velocity profile $U(y,t)$ and polymer
stress components $(\Txx(y,t),\Txy(y,t),T_{yy}(y,t))$, satisfy the equations
\begin{align}
\beta \pdd{U}{y} + \pd{\Txy}{y} &=  -2, \label{315a} \\
\Wi \left(\pd{\Txx}{t} - 2 \Txy\pd{U}{y}\right) + \Yb \Txx
&= \kappa \pdd{\Txx}{y} 
,   \label{bs1}
\\
\Wi \left(\pd{\Txy}{t} - \Tyy\pd{U}{y}\right) + \Yb \Txy
&= (1-\beta)\pd{U}{y} + \kappa \pdd{\Txy}{y} ,
\label{bs2}
\\
\Wi \left(\pd{\Tyy}{t} - \Tyy\pd{U}{y}\right) + \Yb \Tyy
&= \kappa \pdd{\Tyy}{y} ,
\label{bs3}
\end{align}
where
\be
\Yb = \left(1 - \frac{\Bi}{\cTb}  \right)  \Theta(\cTb-\Bi),
\label{321}
\ee
where $\Theta(X)$ is Heaviside's step function and $\cTb$ is now the
base-flow stress invariant.
Equation \eqref{bs3} contains no forcing of the normal stress component
$\Tyy(y,t)$. Therefore, if $\Tyy(y,0)=0$, this component vanishes for all time,
a simplification we adopt in our exploration henceforth.
Consequently, we may ignore \eqref{bs3} and solve \eqref{315a}-\eqref{bs2}
along with
\be
\cTb = \sqrt{\Txx^2+4\Txy^2},
\ee
prescribing initial condition on $\Txx$ and $\Txy$.
In particular, following \citet{evpoiy}, we adopt the
convenient conditions,
\be
[\Txx(y,0),\Txy(y,0)] = [a_0,0]\Bi,
\ee
which introduce no shear stress, but a uniform normal stress
that is a multiple of the yield stress. If $|a_0|<1$, the initial
stress is not sufficient to yield the fluid, but if $|a_0|>1$,
fluid immediately yields.
The introduction of a uniform normal stress is
not meant to mimic any specific flow preparation, but is chosen
for simplicity, and because it generates base flows that differ
from those without any pre-stress, which, as we report
in \citet{evpoiy} and \S\ref{sec:LS}, has impact on the linear
stability.

Numerical solutions of \eqref{315a}-\eqref{bs2} are easily furnished by
introducing centred finite differences
on a uniform grid to approximate derivatives in $y$, then
integrating in time using MATLAB's ode15s solver ({\it i.e.}
using the method of lines).
Note that the initial conditions possess the symmetry $\Txx\to\Txx$
and $\Txy\to-\Txy$ for $y\to-y$, which is maintained for all time
by \eqref{315a}-\eqref{bs2}.

\begin{figure}
\centering
\includegraphics[width=\textwidth]{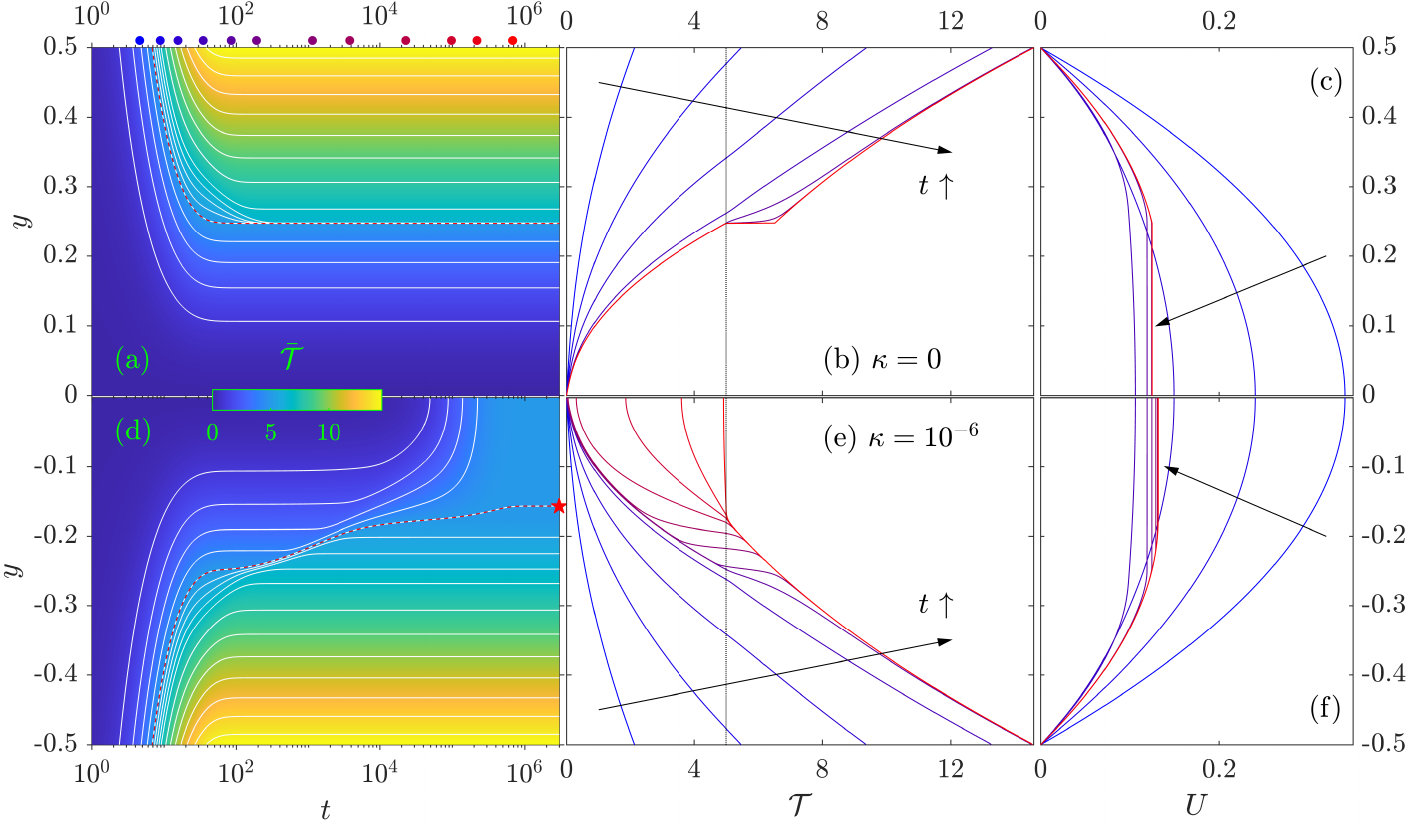}
\caption{Unidirectional flow solutions with
  $(\Wi,\Bn,\beta,a_0)=(10,5,0.5,0)$, for
  (a,b,c) $\kappa=0$ and (d,e,f) $\kappa=10^{-6}$.
  In (a,d), the stress invariant $\cTb$ is plotted as a density
  over the $(t,y)-$plane, with additional
  contours of constant $\cTb$ shown in white; the yield
  surface $\cTb=\Bi$ is indicated by the dashed (red) line).
  Snapshots of $\cTb$ and $U$ are plotted
  in (b,c,e,f) for the times indicated at the top of (a).
  For (a,b,c), the solution is shown in the upper
  half of the channel, and in the lower half for (d,e,f).
  The vertical dotted line in (b,e) indicates the yield condition $\cTb=\Bi$.
  The star in (d) shows the position of the yield surface for the
  diffusive steady state (\S\ref{App:BS}). 
}
\label{fig:basa}
\end{figure}

\begin{figure}
\centering
\includegraphics[width=\textwidth]{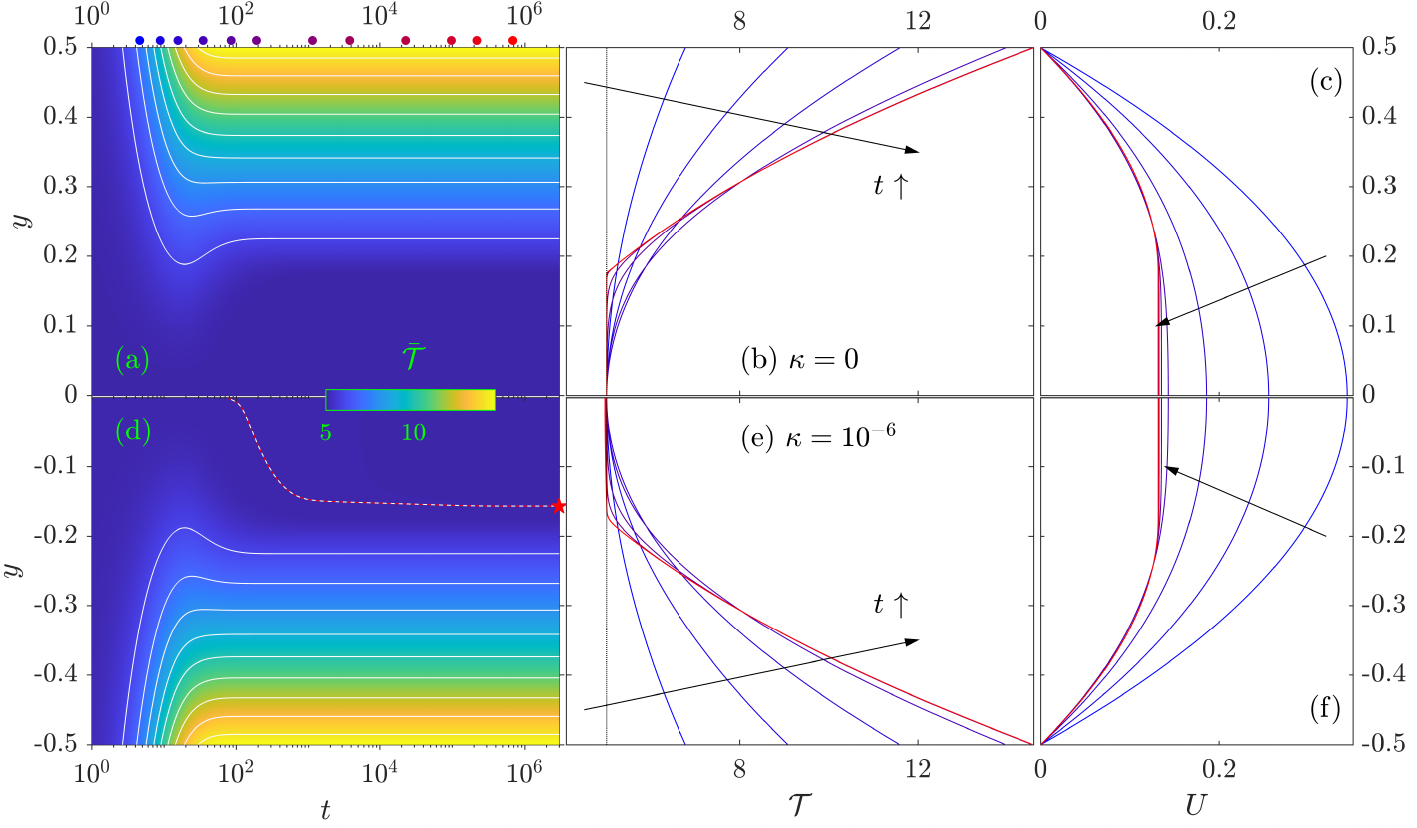}
\caption{A similar set of plots to those in figure \ref{fig:basa},
  but with initial normal stress $a_0=1$.
}
\label{fig:basb}
\end{figure}

Sample solutions for
two representative cases of $a_0$
are illustrated in Figs.~\ref{fig:basa}
and \ref{fig:basb}. In each case, base states are shown without stress
diffusion ($\kappa=0$) and with $\kappa=10^{-6}$.
When $a_0=0$ (Fig.~\ref{fig:basa}), there is no initial polymer stress
and the entire fluid
layer is unyielded. However, stresses build up
through elastic deformation induced by the imposed pressure
gradient until fluid yields at the channel walls. Yield surfaces
$y=\ys(t)$ (where $\cTb=\Bi$)
then migrate towards the channel's centre, but stop before reaching
$y=0$, leaving an unyielded central plug. At least initially,
the base state evolves similarly with or without stress diffusion.
For $\kappa=0$, the stress profile
around the yield surfaces begins to sharpen
for times of $O(10)$. As shown by \cite{evpoiy}, this sharpening
proceeds exponentially quickly, and after infinite time
creates a discontinuity. The discontinuity arises in both 
polymer stress components, $\Txx$ and $\Txy$, although the total
shear stress, $\Txy+\beta \partial U / \partial y\equiv -2y$, remains
continuous. At the stress jumps, the shear rate
therefore also becomes discontinuous, generating breaks in slope
in the final velocity profile (see Fig.~\ref{fig:basa}(c)).

As illustrated in Fig.~\ref{fig:basa}(d,e),
stress diffusion counters the sharpening of the profile at the
yield surfaces. In fact, over longer times, diffusion
causes $\ys(t)$ to continue to migrate further inwards, until
it converges to a final diffusive steady state
(Fig.~\ref{fig:basa}(d)). Simultaneously, diffusion into the plugged
core of the channel gradually raises the stress up to values
much closer to $\Bi$ (Fig.~\ref{fig:basa}(d,e)).
The final diffusive steady states are examined further in
\S\ref{App:BS}.

For the solution in Fig.~\ref{fig:basb} with a pre-stress of $a_0=1$,
the initial fluid layer is yielded throughout. Without
stress diffusion, the stresses and shear rates now remain continuous and
a marginally yielded `pseudo-plug' forms across the centre of the
channel, over which $\cTb$ always slightly exceeds $\Bn$.
The difference between the two final states reached
in Fig.~\ref{fig:basa}(a,b,c) and Fig.~\ref{fig:basb}(a,b,c)
highlight the $x$-independent base flows predicted
by the model equations are not unique when $\kappa=0$
\citep{cheddadi08,hajieghrary26,evpoiy}: the final
state over any unyielded region
depends on the initial stress configuration.
Indeed, other states are possible given different choices of $a_0$
or initial stress profiles
\citep{evpoiy}. Here we focus only on the two cases in
Figs.~\ref{fig:basa} and \ref{fig:basb},
which provide convenient settings in which to explore the flow dynamics.

Diffusion is less significant for the solution
with $a_0=1$ shown in Fig.~\ref{fig:basb}(d,e,f), primarily
because sharp stress gradients no longer develop.
However, diffusion 
permits the stress invariant to fall below $\Bi$ at late times
(see the yield-surface contour in Fig.~\ref{fig:basb}(d)).
A fraction of the pseudo-plug of the non-diffusive problem then becomes
unyielded. The final diffusive adjustments of the two solutions
with $\kappa=10^{-6}$ in Fig.~\ref{fig:basa}(d,e,f)
and Fig.~\ref{fig:basb}(d,e,f)
take place for times of $O(\kappa^{-1})$, and eventually bring
both to the same final state.
In other words, stress diffusion eliminates any indeterminacy of the
plug stress and introduces a unique final steady state.

\subsection{Diffusive steady states}\label{App:BS}

The steady diffusive base flows, with velocity profile $U(y)$ and
stress components $(\Txx(y),\Txy(y))$, satisfy the equations
\begin{align}
\beta \od{U}{y} &= - \Txy - 2y, \label{315} \\
\kappa \odd{\Txx}{y} &= \Yb \Txx - 2\Wi \pd{U}{y} \Txy
,   \label{1.17n}
\\
\kappa \odd{\Txy}{y} &=
  \Yb \Txy - (1-\beta)\od{U}{y} ,
\label{1.18n}
\end{align}
with $\Txx'(\pm\frac12)=\Txy'(\pm\frac12)=0$.
A complication in solving these equations arises when
diffusion allows the base-state
stress to fall below $\Bi$. For the parameter settings chosen in the main
text, this arises when $\kappa$ becomes slightly less than $4\times10^{-4}$
({\it i.e.}, for $\kappa\gtrsim 4\times10^{-4}$, there is no plug).
Once a plug forms by diffusion,
the solution over that region is given by $\Txy=-2y$, with
$\Txx$ taking the constant value found at the yield surface $y=\pm y_y$.
That position is not known, however, and must be found by solving
\eqref{1.18n} subject to the boundary and matching conditions,
$U(\pm\frac12)=\Txx'(\pm\frac12)=\Txy'(\pm\frac12)=0$,
$U'(y_y)=0$, $\cTb(y_y)=\Bi$ and $\Txx'(y_y)=0$. The yield position
then follows as an eigenvalue.

\begin{figure}[t!]
\centering
\includegraphics[width=\textwidth]{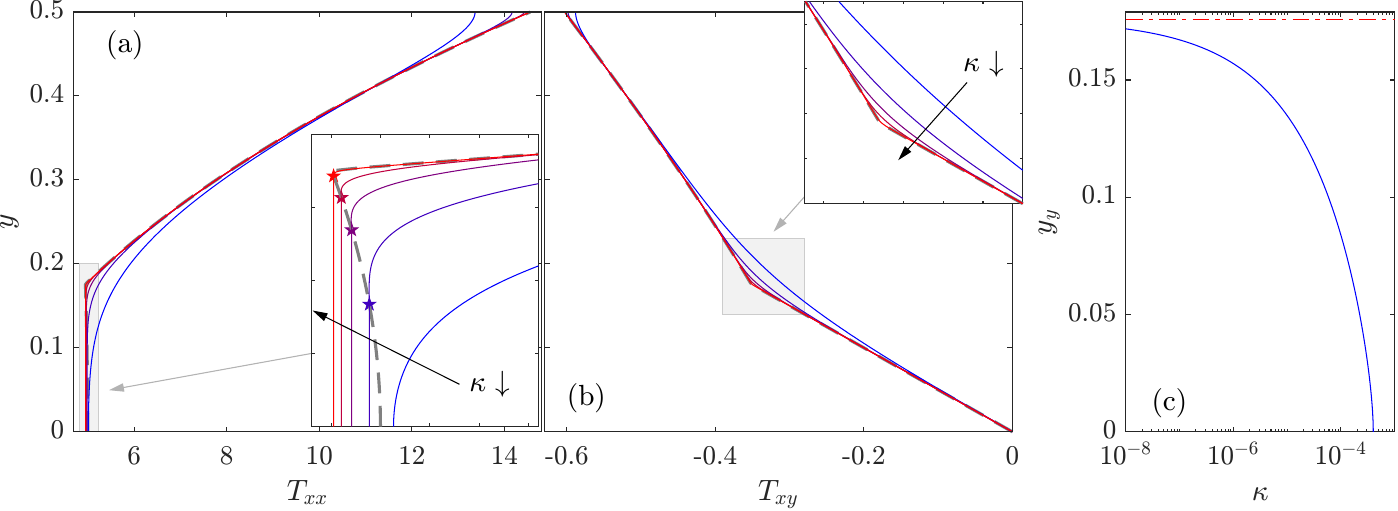}
\caption{Diffusive steady states for
  $(\Wi,\Bn,\beta)=(10,5,0.5)$ and various values of $\kappa$.
  In (a) and (b), we plot the stress components $\Txx(y)$ and $\Txy(y)$
  (respectively) for $\kappa=10^{-j}$, $j=\{3,4,5,6,8\}$
  (colour-coded from blue to red, with $\kappa$ decreasing as indicated
  by the arrows). The insets show magnifications near the yield
  surfaces, which are indicated by filled circles in (a).
  The dependence of the position of the yield surface $y_y$
  on $\kappa$ is shown in (c). The dashed lines in (a,b) show
  the pseudo-plug solution with $\kappa=0$, and the dot-dashed
  line in (c) indicates the corresponding yield position.
}
\label{fig:deqit}
\end{figure}

Figure \ref{fig:deqit}, shows diffusive steady profiles profiles
computed in this manner for a range of stress diffusivities
and the same remaining parameter setting as in Figs.~\ref{fig:basa}
and \ref{fig:basb} (the steady-state
yield surface for $\kappa=10^{-6}$ is also indicated
in panel (d) of those figures). In addition to smoothing the stress
profile near $y=\ys$, less significant
boundary layers also appear against the channel
walls in order to satisfy the no-flux stress conditions applying there.
Note that the diffusive steady state does not converge to the
pseudo-plug of the non-diffusive solution: the latter is
shown by dashed lines in Fig.~\ref{fig:deqit}, and is given
(implicitly)
by
\be
\begin{aligned}
  1 - \frac{\Bi}{\sqrt{\Txx^2+4\Txy^2}}
  &= -\frac{(1-\beta)(\Txy+2y)}{\beta \Txy} ,\\
  \Txx &= \frac{2\Wi \Txy^2}{(1-\beta)},
\end{aligned}
\label{2.17}
\ee
for $\ys<|y|<\frac12$, and
\be
\Txx = \sqrt{\Bi^2-16y^2} \quad {\rm and} \quad
\Txy=-2y,
\ee
for $\ys>|y|$, where
\be
\ys = \frac{(1-\beta)\sqrt2}{4\Wi}
  \left[ \sqrt{\frac{\Wi^2\Bi^2}{(1-\beta)^2}+1} - 1 \right]^{1/2}
\ee
({\it cf.} \citet{evpoiy}).
By contrast, the plug of the diffusive steady state has a constant
normal stress $\Txx$ (see the inset to Fig.~\ref{fig:deqit}(a)).
Nevetheless, for $\kappa\to0$,
the yield position of the diffusive steady state
does converge to that of the $\kappa=0$ pseudo-plug,
as seen in Fig.~\ref{fig:deqit}(c).

\subsection{Linear stability}\label{sec:LS}


Stress discontinuities
in non-diffusive base flows pose a significant challenge to conducting
linear stability analysis \citep{evpoiy}. 
The final steady states reached with diffusion are continuous, however,
permitting their stability to be analysed in detail.
For this task, and continuing to refer to the steady diffusive base state
as $(\Txx,\Txy,U)$,
we first set
\be
\begin{aligned}
  p&=\check{p}(y)e^{ikx+\sigma t},\\
  [\txx,\txy] &= [\Txx(y),\Txy(y)] +
   [\ctxx(y),\ctxy(y)]e^{ikx +\sigma t},\\
  \tyy &= \ctyy(y)e^{ikx},\\
[u,v] &= [U(y),0] + 
   \left[-\psi'(y),ik\psi(y)\right]e^{ikx+\sigma} ,\qquad\qquad
   \label{2.19n}
\end{aligned}
\ee
where $k$ is the wavenumber, $\sigma=\sigma_r+i\sigma_i$ the (complex)
growth rate, and
the prime signifies differentiation
with respect to $y$.
Then, after a linearization of \eqref{conteq}-\eqref{1.13}
and the elimination of $\check{p}$, we find
\begin{align}
\beta (\psi'''' - 2k^2 \psi'' + k^4\psi) 
&= ik(\ctxx'-\ctyy') + \ctxy'' + k^2 \ctxy ,
\label{1.16} \\
  \cD \ctxx  + \frac{\Txx \Yc}{\Wi} -
  2 U' \ctxy
  &=  - \frac{2ik}{\Wi} (1-\beta) \psi'
  - (ik \Txx' \psi 
  + 2ik \Txx \psi' 
  + 2  \Txy \psi'' )
  \label{1.17}
    , \\
\cD \ctxy + \frac{\Txy \Yc}{\Wi}
  -  U' \ctyy   
  &= - \frac{1}{\Wi}(1-\beta) (\psi'' + k^2\psi)
- (ik \Txy' + k^2\Txx) \psi
  \label{1.18}
  , \\
 \cD \ctyy 
  &= \frac{2ik}{\Wi}(1-\beta) \psi' 
  - 2 k^2\Txy \psi,
  \label{1.19}
\end{align}
where
\be
\cD = \sigma + ikU + \frac{\Yb}{\Wi} - \frac{\kappa}{\Wi}
  \left( \pdd{}{y} - k^2 \right),
\ee
the base-state switch in \eqref{321} is again written as $\Yb$,
and
\be
\Yc = \frac{\Bi}{\cT^3} \left[\Txx (\ctxx-\ctyy)
+ 4 \Txy \ctxy\right] \Theta(\cTb-\Bi)
\ee
denotes the perturbation to the switch.
The boundary conditions on the walls become
\be
\psi(\pm\half)=\psi'(\pm\half)=
\ctxx'(\pm\half)=
\ctxy'(\pm\half)=
\ctyy'(\pm\half)=
0 .
\label{3.21}
\ee
We solve the system \eqref{1.16}-\eqref{3.21} in two ways:
first, given the base state solution on a uniform spatial
grid in $y$, we replace derivative by centred finite differences
and break the system down to a matrix eigenvalue problem
for the perturbed stress components. Second, we solve the system
as a boundary-value problem using MATLAB's bvp4c,
armed with a trial based on a
finite-difference solution or another
boundary-value solution with different parameter settings.

\begin{figure}[t!]
\centering
\includegraphics[width=0.99\textwidth]{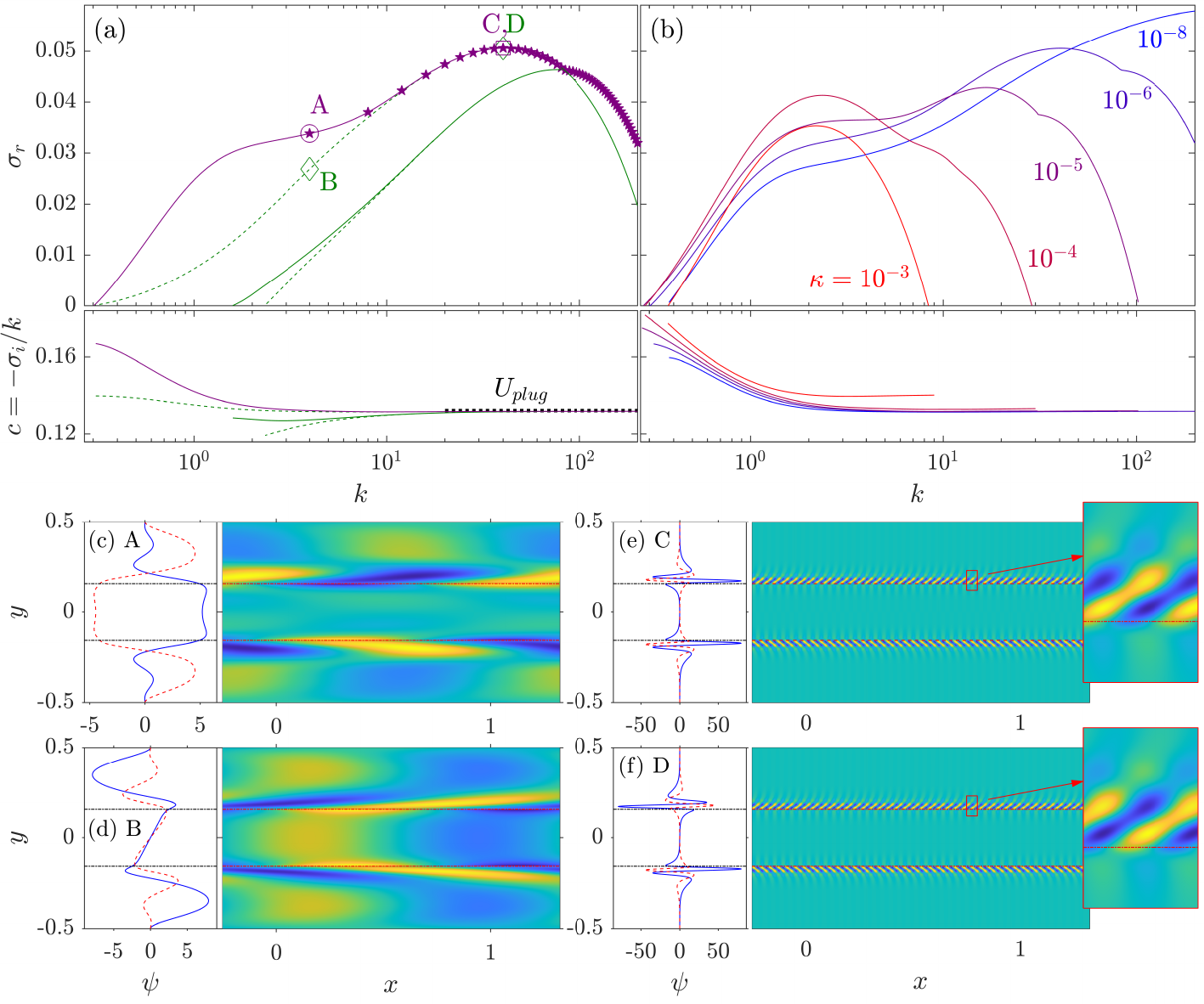}
\caption{ Growth rates and phase speeds of the unstable normal modes
  against wavenumber $k$,
  for diffusive steady states with
  $(\Wi,\Bn,\beta,a_0)=(10,5,0.5,0.5)$ and 
  (a) $\kappa=10^{-6}$ or (b) varying stress diffusivities
  (for the values of $\kappa$ indicated).
  The stars in (a) mark the growth rates for modes in a domain
  of length $L=\frac12\pi$ and the two most unstable even and odd modes are
  both plotted, with the odd modes (in $\psi(y)$)
  shown by dashed lines. In (b), only
  the most unstable modes at each wavenumber are plotted.
  The dotted line indicates the plug speed $U_{plug}$. The
  modal eigenfunction for $\psi$ (solid for the real part
  and dashed for the imaginary part) and the
  spatial patterns of the normal stress perturbation
  $\ctxx$ (as a density over the $(x,y)-$plane)
  are shown in (c) for the modes
  identified as A, B, C and D. A and B correspond to the gravest modes
  with $L=\frac12\pi$; C and D to the most unstable ones over all
  the higher $k$ values.
  The dot-dashed lines indicate the
  yield surface of the base state, and the overlaid plots
  for C and D show local magnifications.
  The linear eigenmodes are normalized so that
  $\int_0^{1/2} \psi \; \di y = 1$. 
}
\label{fig:bas66}
\end{figure}

Fig.~\ref{fig:bas66} shows sample results for the base states 
from Fig.~\ref{fig:deqit}.
Modal growth rates $\sigma_r$, phase speeds
$-\sigma_i/k$ and spatial patterns (in the perturbed
extensional stress component $\ctxx$) are displayed
for varying wavenumber $k$ and different choices for $\kappa$.
For the lowest values of $\kappa$, growth rates are highest
at the largest wavenumbers, as found
previously for $\kappa=0$ \citep{evpoiy}.
Simultaneously, the phase speed of the modes converges to the
speed of the plug $U_{plug}$ and the modal eigenfunctions become
localized to the vicinity of the yield surface
(see Fig.~\ref{fig:bas66}(e,f)).

Stronger stress diffusion reduces the growth rates for high $k$,
introducing a preferred wavenumber.
Lower wavenumbers, instead, can be destabilized by polymer diffusion,
a result stemming mostly from the diffusive adjustment to the
yield surfaces of the base state. This destabilization is not
equivalent to the polymer-diffusion instability found for
viscoelastic fluids \citep{beneitez23,beneitez25}, which
arises at the walls and was not observed in our elasto-viscoplastic problem.
Indeed, the instability found here requires no polymer diffusion
whatsoever \citep{evpoiy}.

Note that there is a wide range of unstable modes at the
  parameter settings of Fig.~\ref{fig:bas66}, in terms of both the
  streamwise wavenumber $k$ and the
  cross-stream normal-mode structure
  (figure \ref{fig:bas66}(a)). The most unstable modes
  at each $k$ correspond to perturbations in $u$ or $\txx$
  that are anti-symmetrical about the midline of the channel
  (symmetrical modes in $\psi$). However,
  the even and odd modes reach similar growth rates at higher
  $k$, a consequence of the localization of the modes to the
  yield surfaces (see Fig.~\ref{fig:bas66}(c)).

For the stress diffusivities shown in Fig.~\ref{fig:bas66}, 
the largest growth rates occur across a relatively wide
range of wavenumbers and spatial mode structure.
Consequently, linear growth
in the numerical simulations reported next
is expected to be relatively insensitive to the length of the domain
and the manner in which $x$-dependent perturbations are initialized.
This observation leads us to select a
domain length of $L=\tfrac{\pi}{2}$ 
and seed the simulations with small-amplitude wavy
perturbations to $\txx$ that are uniform in $y$ and
have wavenumber $k=4$. The initial conditions therefore
have a simple spatial structure and possess the
longest wavelength in the domain.
However, a relatively wide range of unstable modes is accommodated
in the domain, thereby
striking a balance between a rich linear dynamics 
and numerical resolution.

Beyond the effect of the polymer diffusivity, one might wonder
about the impact of the other model parameters on the linear
stability problem. This question was addressed by \citet{evpoiy}
who showed that the growth rate of the most unstable, short-wavelength
modes with $\kappa=0$ took the form,
\be
\sigma_r = \Wi^{-1} \Sigma\left(X ;\beta\right)
,
\qquad X = \Wi\Bi
= \frac{2 \tau_y \lambda}{\mu}
,
\label{e:gr}
\ee
for some function $\Sigma(X,\beta)$, with a relatively weak
dependence on the second argument $\beta$.
Note that \citet{evpoiy} employ a different scaling of time for the problem,
resulting in the absence of the factor of $\Wi^{-1}$ in their formulae
(and employ different dimensionless parameters, writing $X$ in a different
way). In addition, \eqref{e:gr} implies a dimensional growth rate
of $\mathcal{H}^{-1}\cU \sigma_r \equiv \lambda^{-1} \Sigma(X ;\beta)$.
The timescale of instability is therefore set by the polymer relaxation time,
with the growth rate controlled primarily by the combination
$\tau_y\lambda/\mu$, which is proportional to the ``yield strain''
of the elasto-viscoplastic material, $\tau_Y/G=\tau_Y\lambda/\mu_p$, where
$G=\mu_p/\lambda$ is the elastic modulus
\citep{moschopoulos2023dynamics,varch,esposito25}.
Although polymer diffusion eliminates the high-wavenumber instability,
as long as $\kappa$ remains sufficiently small one expects the
parameter dependence in \eqref{e:gr} to remain key.
 
\section{Nonlinear dynamics}\label{sec:NS}

\subsection{Solution strategy}

We solve \eqref{conteq}-\eqref{bcs} in a periodic domain, $0\leq x\leq L$,
taking $L=\tfrac{\pi}{2}$ and using the open-source spectral solver
Dedalus \citep{burns2020dedalus}.
Further details are provided
in Appendix \ref{app:num}, along with a report of tests of the
fidelity of the numerical scheme. 
Simulations
are run until a fixed time, which we
take to be at least $t=1000$, but as much as $t=6000$.
We set $\beta=0.5$ throughout, representing moderate solvent viscosity,
and mostly use diffusion coefficients over the range
$10^{-5}\leq \kappa\leq 10^{-6}$ to avoid substantial stress diffusion
whilst ensuring adequate spatial resolution. We discuss the impact of varying $\kappa$ in more detail in Appendix \ref{app:num}.
As initial conditions, we adopt the same initial stresses
as used to explore the base states, but including an additional,
small-amplitude
wavy perturbation with a specific wavenumber:
\be
\left[\txx(x,y,0),\txy(x,y,0),\tyy(x,y,0)\right]=
\Bi \left[a_0+10^{-3}\sin\frac{2\pi x}{L},0,0\right].
\label{IC1}
\ee
Note that the initial condition \eqref{IC1} possesses symmetry about the centreline of the
channel, $y=0$, a feature that is maintained for all time by solutions
to \eqref{conteq}-\eqref{bcs} 
(the streamwise velocity $u$ and stress components $\txx$ and $\tyy$
are even in $y$, whereas $v$ and the shear stress $\txy$ are odd).
In the numerical simulations, the solutions possess
this symmetry up to times of $O(10^3)$, but then
round-off errors eventually trigger a loss of
the symmetry about $y=0$.

  \subsection{Linear growth and initial saturation}\label{sec:phas1}

\begin{figure}[t!]
\centering
\includegraphics[width=.99\textwidth]{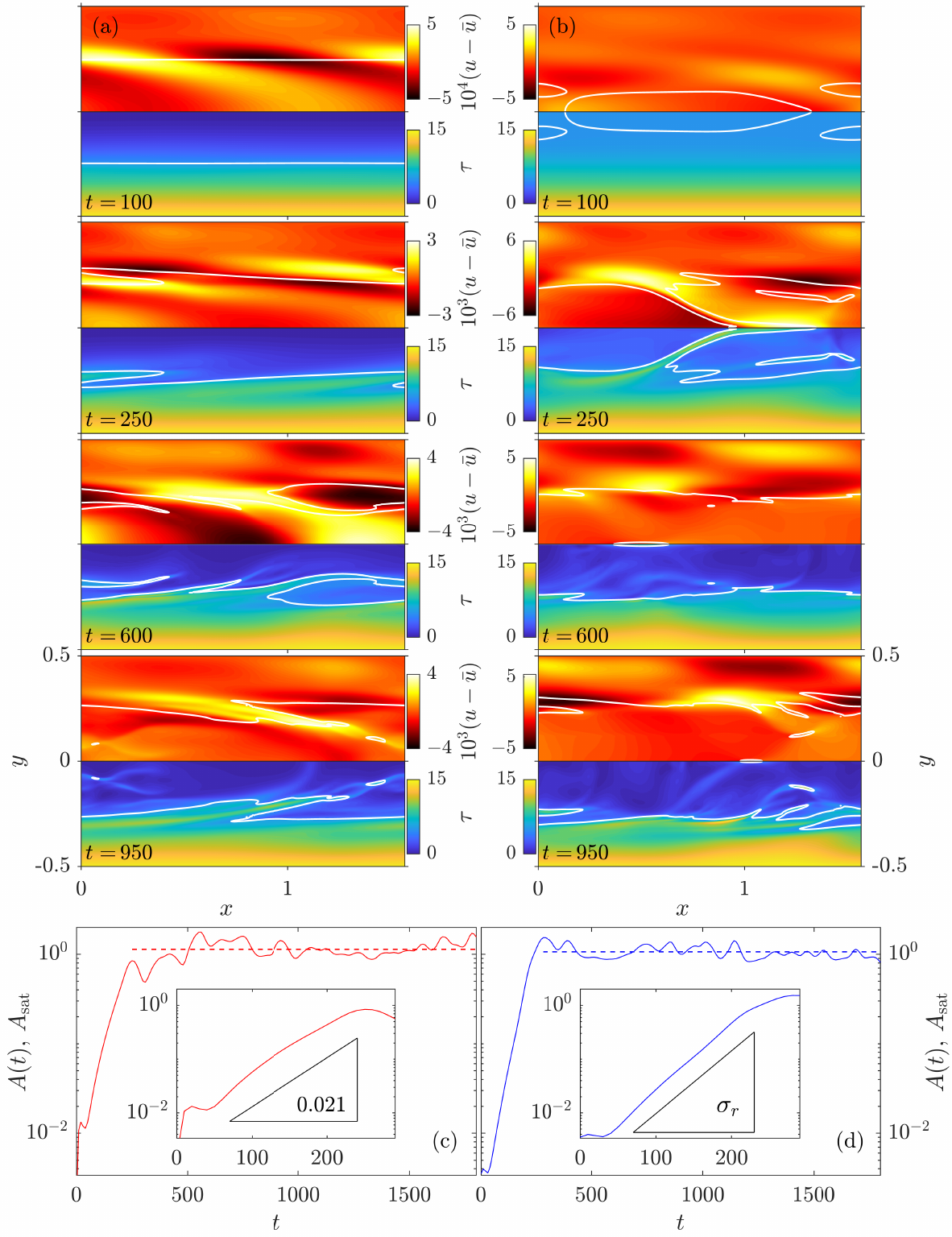}
\caption{Snapshots of $u-\bar{u}$ and $\tau$
  (at the times indicated and shown over the upper or lower halves
  of the channel, respectively),
  from numerical simulations with (a) $a_0=0$ and (b) $a_0=1$, for
  $\{\Wi,\Bn,\beta,\kappa,L\}=\{10,5,\tfrac{1}{2},10^{-6},\tfrac{\pi}{2}\}$.
  The full solutions remain symmetrical about the centreline at these times.
  Solid white lines indicate the yield surfaces where $\tau=\Bn$.
  (c,d) Time series of the fluctuation amplitude $A(t)$ defined in
  \eqref{Adef} for the two simulations;
  dashed lines show the mean value $\Aend$ in \eqref{Aend}.
  The insets show magnifications of the linear growth stage; the triangles
  indicate exponential growth at a rate $0.021$ in (a)
  (the value of \eqref{grapp} for this case,
  as plotted in Fig.~\ref{fig:sim2}), and at
  the linear growth rate $\sigma_r$ of the mode labelled B in
  Fig.~\ref{fig:bas66}.
  (Videos of the two solutions are provided as Supplementary
  Materials.) 
 }
\label{fig:sim}
  \end{figure}

  To illustrate the nonlinear dynamics of the system, we use
  the two examples whose base states are shown in Figs.~\ref{fig:basa}
  and \ref{fig:basb} (with 
  $\{\Wi,\Bn,\beta,\kappa,L\}=\{10,5,\tfrac{1}{2},10^{-6},\tfrac{\pi}{2}\}$
  and $a_0=0$ or $a_0=1$),
  breaking the evolution down into the earlier-time symmetrical phase
  and the later desymmetrized one.
  Fig.~\ref{fig:sim} presents results for the first, symmetrical stage.
  In the upper rows of this figure, the symmetry
  about the centreline $y=0$ is exploited
to compactly display snapshots of the evolving stress invariant
$\tau$ (for $y<0$) and streamwise velocity perturbations
$u-\bar{u}$ (in $y>0$), where 
\be
\bar{f} = L^{-1}\int_0^L f\; \mathrm{d}x
\ee
denotes a streamwise average.

Focusing first on the case with $a_0=0$, 
evolution begins with a short transient adjustment
from the initial configuration towards a state resembling the corresponding
base-flow solution. This adjustment lasts about about ten time units
and creates a plug spanning the centre of the channel
However, small-amplitude
perturbations are present
with wavelength $\tfrac{\pi}{2}$, seeded by the initial condition in
$\txx$ (see Fig.~\ref{fig:sim}(a), at $t=100$). We gauge 
the strength of those perturbations by defining an amplitude,
\begin{equation}
 A(t) = \left[\frac{1}{L}\int_0^{L}\int_{-\frac{1}{2}}^{\frac{1}{2}}(\tau-\bar{\tau})^2\,\mathrm{d}x\mathrm{y}\right]^{1/2},
  \label{Adef}
\end{equation}
whose time series is plotted in Fig.~\ref{fig:sim}(c). The amplitude
grows somewhat exponentially in time, until about $t=200$, at which
point $A(t)$ reaches order-one values and 
perturbations become nonlinear (see the inset to Fig.~\ref{fig:sim}(c)). 
Prior to saturation, the linear growth is not perfectly exponential because
the perturbations amplify on an evolving base state whose
stress gradients also sharpen exponentially quickly
at the yield surfaces (see \citet{evpoiy}).

The saturation of $A(t)$ corresponds to the emergence of a nonlinear wave 
that is localized around, and
significantly deforms, the yield surfaces (Fig.~\ref{fig:sim}(a), $t=250$).
The nonlinear wave does not become steady, however, but continues
to evolve, with $A(t)$ then
oscillating irregularly about an approximately constant level
shown by a dashed line in Fig.~\ref{fig:sim}(c)).
This level, $\Aend$, is given by the temporal average,
\begin{equation}
  \Aend = \frac{1}{t_\mathrm{end}-t_\mathrm{peak}}\int^{t_\mathrm{end}}_{t_\mathrm{peak}}A(t)\,\mathrm{d}t,
  \label{Aend}
\end{equation}
where $t=t_{\mathrm{peak}}>150$ is the time at which $A$ has its first maximum
greater than $0.002$
(to eliminate any maxima during early-time transients),
and $t_{\rm end}$ is the end of the relevant time interval
(in this case $t=1800$).
Simultaneously, shorter wavelengths emerge in the spatial
structure, as illustrated by the appearance of multiple
streaks in the stress field and fingers of unyielded fluid
protruding into the central plug ({\it cf.} the snapshot at $t=600$).
Occasionally, stresses fall back below
$B$ along those fingers, isolating unyielded pockets within the plug
({\it e.g.} the final snapshot for $t=950$). 

With a pre-stress that initially yields the fluid
($a_0=1$), the initial transient is again followed by linear growth
(Fig.~\ref{fig:sim}(b,d)). Now, amplification is more
closely exponential, as shown in Fig.~\ref{fig:sim}(d) where
the growth in the simulation is compared
with the predictions of the linear theory
of \S \ref{sec:LS} for the odd normal mode in $\psi(y)$
with a wavenumber of $k=4$ ({\it i.e.} mode B in Fig.~\ref{fig:bas66}).
Nevertheless, although the stress
remains close to $\Bn$ across the centre of the channel
({\it cf.} Fig.~\ref{fig:basb}), several spatially structured plugs
form as $\tau$ locally drops below the yield stress there, which
expand as the perturbations grow with time. The plug structure
in the simulation therefore evolves differently than for the
corresponding base state with a pseudo-plug, implying that
the growth rate predicted by linear theory should not match completely.

For $t=250$, $A(t)$ again saturates; see Fig.~\ref{fig:sim}(d).
At this stage, the stress field over the core of the channel is different to
the $a_0=0$ simulation, with thin bands of strongly yielded fluid
protruding to the centreline, fracturing the emerging plug there. 
By $t=600$, the entire centre of the channel becomes plugged up
(see Fig.~\ref{fig:sim}(b)), 
except for a few small pockets of unyielded fluid.
Multiple streaks again appear in the stress field,
complicating the yield surfaces and highlighting the excitation of
shorter wavelengths (see the snaphost at $t=950$). During the later stages,
the dynamics observed in two simulations becomes more similar.

\begin{figure*}[t!]
\centering
\includegraphics[width=\textwidth]{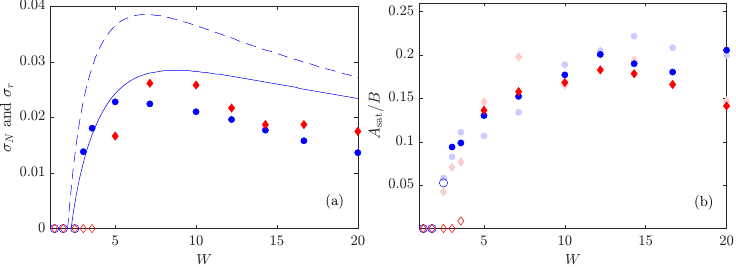}
\caption{Data from two suites of simulations
  with varying $\Wi$ and
$\left(\Bn,\beta,\kappa,L\right)=\left(\tfrac{1}{2}\Wi,\tfrac{1}{2},10^{-5},\tfrac{\pi}{2}\right)$,
  with $a_0=0$ (blue circles) or $a_0=1$ (red diamonds). 
  (a) Approximate early-time growth rate $\sigma_N$ from \eqref{grapp}.
  The lines show the actual linear growth rates of the even (solid) and
  odd (dashed) modes in $\txx$. 
  (b) Scaled final amplitude $\Aend/\Bi$, as defined in \eqref{Aend}, for the same simulations as in (a) (darker symbols),
  with $t_{\rm end}=10^3$.
  The fainter symbols are equivalent data from another suite of simulations in which the initial conditions \eqref{IC1} are modified: instead of applying the single sine wave perturbation to $\txx$, low-amplitude random noise is applied to $\txx$ initially. 
  In both (a) and (b), open symbols correspond to simulations that
  did not reach a peak in $A$
  before $t=1000$; in these cases $\Aend$ is replaced by $A(10^3)$ in (b). 
}
\label{fig:sim2}
\end{figure*}

Fig.~\ref{fig:sim2} illustrates the impact of elasticity on
the first phase of evolution,
collecting together data for two diagnostics from suites of simulations
with varying $\Wi$. 
In addition to fixing
$a_0=0$ or 1 and the same values of $\beta$ and $L$ in these computations,
we set $B=\frac12 \Wi$, which leads to unyielded plug regions
of roughly comparable size.
We adopt the slightly higher diffusivity
$\kappa=10^{-5}$, permitting computations at higher $\Wi$.
That said, we confirm in Appendix \ref{app:num} that
the results for $\Wi=10$ and $0\leq t\leq 1000$
are similar to those in Fig.~\ref{fig:sim}(b).

The first diagnostic shown in
Fig.~\ref{fig:sim2} is a rough measure of the growth
rate of instability, defined by
\be
\sigma_N = 
\frac{1}{t_\mathrm{peak}}\log \frac{A(t_\mathrm{peak})}{A(0)}.
\label{grapp}
\ee
This diagnostic would equal the linear growth rate
were $A(t)$ to follow an exponential linear
amplification for $0<t<t_{\rm peak}$.
In the simulations, however, the amplification
is obscured by initial transients, the nonlinear saturation process,
and the fact that linear growth is not purely exponential
(Fig.~\ref{fig:sim}(c,d)).
Despite this, the measurements of \eqref{grapp}
show qualitative agreement with the computed linear growth
rates for such base states (also included in Fig.~\ref{fig:sim2}(a)).
In particular, the simulations
commencing from a pseudo-plug ($a_0=1$)
reproduce the onset of linear instability near $\Wi=2$,
a maximum in growth near $\Wi=6$, and a final decline
of the growth rate for larger $\Wi$.
The simulations with zero initial polymer stress ($a_0=0$)
display a slightly higher 
critical value of $\Wi$ for onset, closer to $\Wi=4$, but the
growth rates are otherwise similar to those from the simulations
with $a_0=1$.

The second diagnostic in Fig.~\ref{fig:sim2}(b,c) is
the scaled mean fluctuation amplitude $\Aend/\Bi$ (with $t_{\rm end}=10^3$).
Beyond the slightly different onsets,
this diagnostic varies relatively weakly with $\Wi$
and is comparable between the simulations with $a_0=0$ (b) and $a_0=1$ (c). 
The plots of $\Aend/\Bi$ in Fig.~\ref{fig:sim2}(b,c) also includes data from
two additional sets of simulations using different initial conditions
to \eqref{IC1}. In these cases, rather that kicking the simulation
off from a single wavenumber as in \eqref{IC1}, we applied low-amplitude random noise to $\txx$ to excite a broad range of wavenumbers. To avoid any numerical issues, a high-wavenumber filter was applied to the random noise that 
retained $20\%$ of the Fourier and Chebyshev modes.
Otherwise, the simulations are set up in the same way,
with $a_0=0$ or $a_0=1$ 
still applied to $\txx$ along with the random noise. Above onset, the resulting values of $\Aend/\Bi$ are broadly
similar between all the simulations, suggesting that the typical level of complexity of the
nonlinear states triggered by linear instability does not depend
sensitively on initial conditions. 

\subsection{Longer-term dynamics}
\label{sec:diff}

  \begin{figure}[t!]
\centering
\includegraphics[width=.97\textwidth]{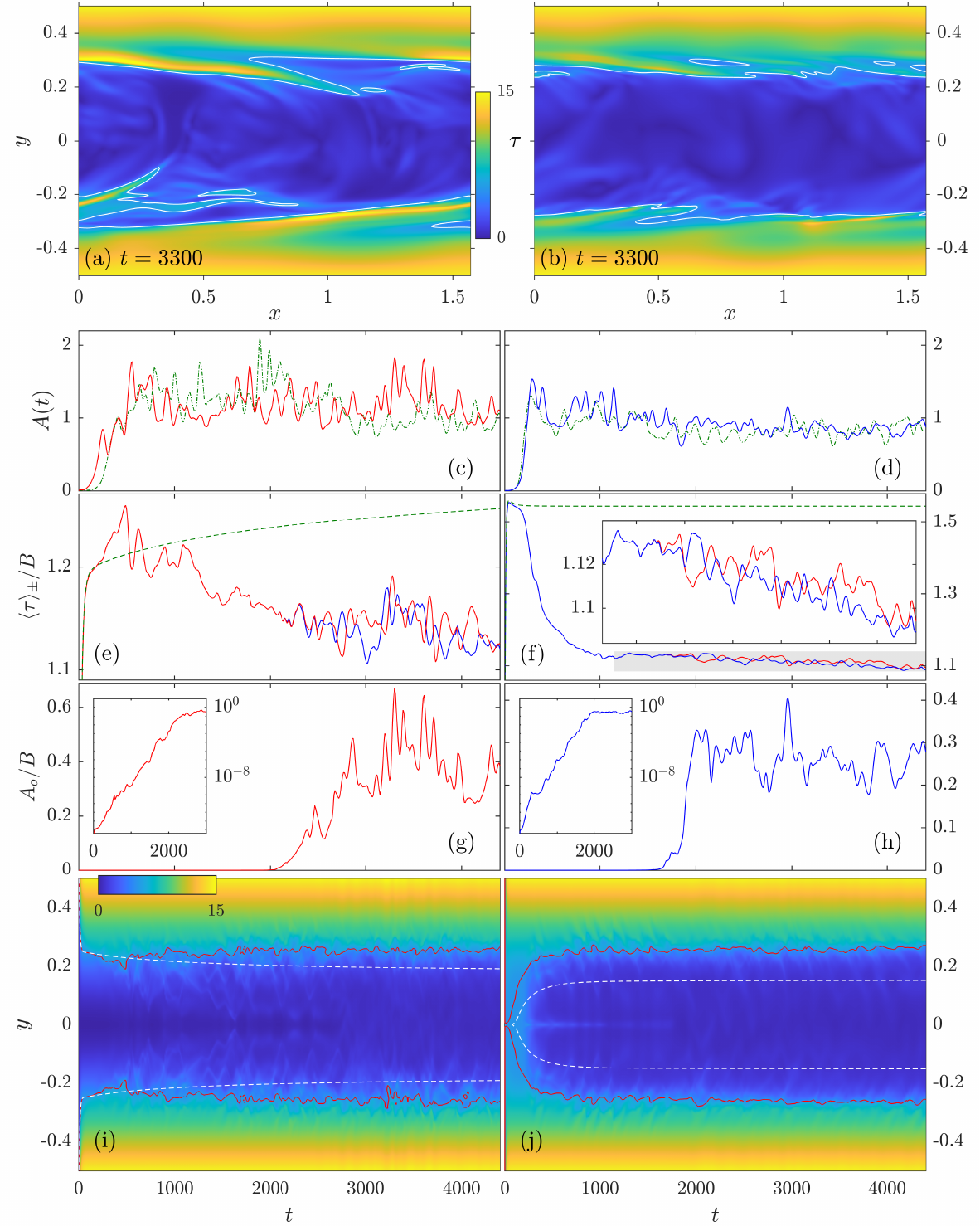}
\caption{
  Longer-time dynamics of the simulations shown in Fig.~\ref{fig:sim},
  with the example with $a_0=0$ on the left, and that with $a_0=1$
  on the right. In (a,b), we present the a snapshot of the stress
  invariant $\tau$ at $t=3300$ as a density over the $(x,y)-$plane;
  solid white lines indicate the yield surfaces.
  Times series are presented below, showing (c,d) $A(t)$ in
  \eqref{Adef}, (e,f) the scaled net stress over the upper and lower
  halves of the channel $\langle\tau\rangle_\pm/\Bi$, and
  (g,h) the scaled root-mean-square odd stress component
  $A_o(t)/\Bi$ from \eqref{Aodef}.
  The dot-dashed lines in (c,d) show equivalent results for $A(t)$
  from simulations with initial conditions in which
  multiple wavelengths are included, not just $k=4$
  (see the main text for further details).
  The dashed lines in (e,f) show the average stress of the
  uni-directional base states.
  The inset in (f) shows a magnification of the shaded region,
  and those in (g,h) replot the $t<3000$ data in the main panels on a
  logarithmic scale.
  In the bottom row (i,j) we show the $x-$averaged stress invariant
  $\overline{\tau}$ as a density on the $(t,x)-$plane. The red contours
  show the mean yield positions where $\overline{\tau}=\Bi$, and the
  white dashed line
  shows the yield positions of the corresponding base states.
 }
\label{fig:mycodi}
  \end{figure}

Fig.~\ref{fig:mycodi} shows the later-time dynamics for the
two simulations in Fig.~\ref{fig:sim}.
Near $t=2000$, both solutions lose symmetry about the centerline
owing to the amplification of round-off errors.
Mildly asymmetric stress profiles then result, as illustrated
by the snapshots of Fig.~\ref{fig:mycodi}(a,b) and the space-time
density plots of the $x-$averaged stress $\overline{\tau}$ in
Fig.~\ref{fig:mycodi}(i,j). The de-symmetrization of the solution
is quantified further in Fig.~\ref{fig:mycodi}(e-h)),
which plot time series of the stress averaged over the
upper and lower halves of the channel,
\be
\langle\tau\rangle_\pm = \pm\int_0^{\pm1/2} \overline{\tau}\, \di y,
\ee
and the root-mean-square odd stress component, defined as
\be
A_o(t) = \sqrt{ \langle [\tau(x,y,t) - \tau(x,-y,t)]^2 \rangle}.
\label{Aodef}
\ee
Here, the angular bracket denotes an average over the channel:
\be
\langle ... \rangle = \frac{1}{L}\int_0^L \int_{-1/2}^{1/2} (...) \di x\; \di y
.
\ee

Note that the mean stresses plotted in Fig.~\ref{fig:mycodi}(e,f)
decrease beyond the saturation of the instability. By contrast,
the mean stress of the corresponding base states, which is also plotted
by the dashed lines in the figure, increases at later times as a result
of stress diffusion. In other words, at finite amplitude
the instability lowers the net stress, countering the effect of
diffusion. The reduction arises mostly over
the centre of the channel where the nonlinear dynamics
acts to maintain relatively low stress levels
(see Fig.~\ref{fig:mycodi}(i,j)). By contrast,
diffusion drives the stress there towards the yield stress
(see Figs.~\ref{fig:basa} and \ref{fig:basb}).
Nevertheless, as we show in Appendix \ref{app:num},
stress diffusion does inevitably impact the dynamics
at times beyond those shown in Fig.~\ref{fig:mycodi},
leading to a run-down of the spatio-temporal complexity.
Consequently, for our main discussion, we
focus on times $t\lesssim4000$, for which the effects of diffusion
appear minimal.

\begin{figure*}[t!]
\centering
\includegraphics[width=\textwidth]{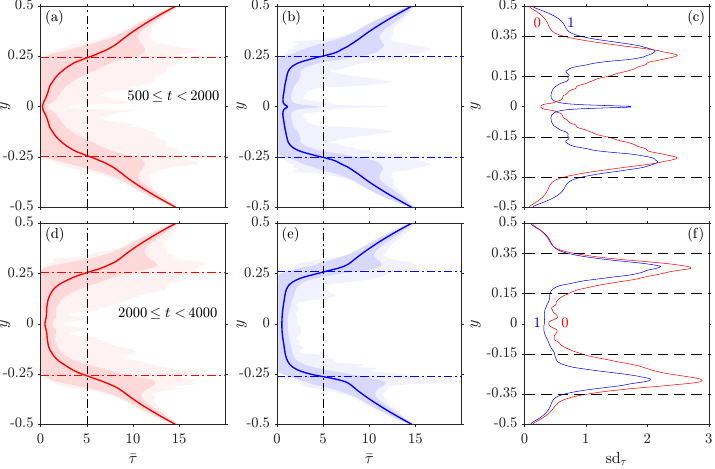}
\caption{(a,b,d,e) Mean streamwise stress $\overline\tau$ (solid lines) for the simulations
  with (a,d) $a_0=0$ and (b,e) $a_0=1$ from Fig.~\ref{fig:mycodi},
  averaged over the time
  intervals of (a,b) $500\leq t<2000$ and (d,e) $2000\leq t<4000$. The
  darker shaded areas span two standard deviations
  from the mean in the $\tau$
  data across all $x$ and the same ranges in $t$;
  the lighter shaded areas indicate the corresponding maximum and minimum values
  of $\tau$. The means and standard deviations are calculated from simulation
  data sets saved at intervals of $10$ time units. Horizontal dot-dashed lines are mean yield surfaces, i.e., the intersections of $\bar{\tau}$ with $\Bi$ (vertical dot-dashed line). (c,f) Standard deviation of the $\tau$ data from the same simulations across the same time intervals.
  The lines corresponding to $a_0=0$ and $a_0=1$ are colour-coded and labelled $0$ and $1$, respectively. Dashed lines demarcate the intervals $0.15<|y|<0.35$. 
}
\label{fig:tau}
\end{figure*}

The suppression of the stress over the plug is illustrated further
in Fig.~\ref{fig:tau}. Here,
the streamwise mean stress, $\overline{\tau}(y,t)$,
is shown averaged in time 
over the intervals corresponding to either the symmetrical
($500<t<2000$; panels (a,b)) or desymmetrized ($2000<t<4000$; panels (c,d))
nonlinear states. Aside from the remnant of some initial transients in 
the former,\footnote{
For example, the narrow spike near the centre line in Fig.~\ref{fig:tau}(b,c)
for the simulation with $a_0=1$ arises
from a persistent small pocket of yielded fluid near $y=0$, 
a remnant of the early-time
formation of the central plug (see Fig.~\ref{fig:sim}(b) at $t=600$);
this feature becomes erased at later times (Fig.~\ref{fig:tau}(e,f)). }
the mean stress is similar between the two time phases and
the two simulations. 

More significant differences between the two simulations arise in
the fluctuations about the average stresses, which are also shown
in Fig.~\ref{fig:tau}. In (a,b,d,e),
the darker shading spans two standard deviations
from the mean, whereas the lighter shading indicates the extreme values
reached within the channel over the respective time intervals.
The standard deviations themselves are plotted in Fig.~\ref{fig:tau}(c,f).
The shadings and deviations emphasise how fluctuations become relatively weak
near the channel walls and centre, and are mostly concentrated 
in the vicinity of the yield surfaces.

\begin{figure*}[t!]
\centering
\includegraphics[width=\textwidth]{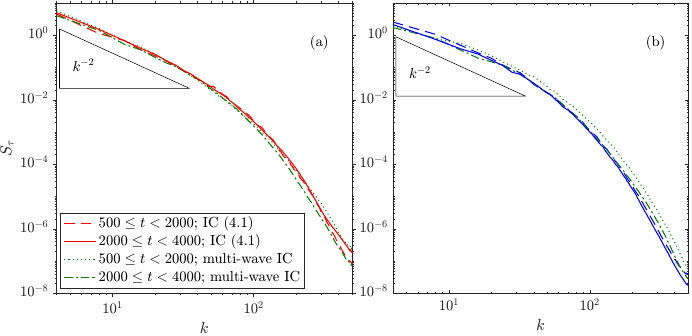}
\caption{Average spectra $S_\tau(k)$ 
  for (a) $a_0=0$ and (b) $a_0=1$, defined as
  in \eqref{specdef}, for the simulations of Fig.~\ref{fig:mycodi}.
  The dashed lines show the early time phase with
  $(t_1,t_2)=(500,2000)$; the solid lines represent the second phase
  for $(t_1,t_2)=(2000,4000)$.
  The triangles show the power-law scaling $k^{-2}$.
  The dotted and dot-dashed lines show the same results, but for the
  simulations with multi-wave initial conditions.
}
\label{fig:spec}
\end{figure*}

Fig.~\ref{fig:tau}(c,f) indicates that the main variations of $\tau$ are contained within the sections $0.15<|y|<0.35$ marked by dashed black lines. To illustrate the broad range of spatial scales excited over these sections, we show streamwise spatial spectra of $\tau$ in Fig.~\ref{fig:spec}, averaged over the two time intervals
and the windows of $y$. That is, for the two simulations, we plot
\begin{equation}
	S_{\tau}(k) = \frac{1}{0.2(t_1-t_0)}\int_{t_0}^{t_1}\left[\int_{-0.15}^{-0.35} |{\cal F}\{\tau\}|^2\,\mathrm{d}y + \int_{0.15}^{0.35} |{\cal F}\{\tau\}|^2\,\mathrm{d}y\right]\mathrm{d}t,\label{specdef}
\end{equation}
where ${\cal F}\{\tau\}$ denotes Fourier transform in $x$, and $[t_0,t_1]$ is the relevant time interval.
The spectra are all similar
and decrease approximately as $k^{-2}$ at low wavenumbers, before
steepening 
at higher wavenumbers. 
Comparable power-law scalings have been reported
for purely elastic turbulence, although the
spectrum of the trace of the conformation tensor is more typically
presented 
\citep[e.g., ][]{lellep24,garg21,steinberg21}. 


Fig.~\ref{fig:spec} also includes results for two more simulations run
with the multi-wave initial conditions described at the end of \S
\ref{sec:phas1}. The time traces of $A(t)$ for these simulations
are also shown in Fig.~\ref{fig:mycodi}(c,d).
Neither $A(t)$ nor the spectra look very
different after changing the initial conditions on the perturbations,
reinforcing our earlier conclusions regarding the saturation level $\Aend$
in Fig.~\ref{fig:sim2}(b). The diagnostics also do not vary appeciably
between the two phases of evolution. Therefore, the complex nonlinear
states reached through linear instability are relatively insensitive to
initial conditions. Fluctuations do appear slightly stronger in the
simulations commencing with no initial polymer stress (compare panels (c)
and (d) in Fig.~\ref{fig:mycodi}, or the spectral levels
between (a) and (b) in Fig.~\ref{fig:spec}), but the intrinsic
variability renders any definitive conclusion difficult.

\section{Discussion}

In this paper,
we have demonstrated that linear instability
triggers a spatio-temporally complicated
flow state in the inertialess channel flow
of an elasto-viscoplastic fluid described by Saramito's Bingham model.
Spatio-temporal complexity is focused to the vicinity of the yield surfaces
of the plug spanning the centre of the channel.
The instability arises at moderate solvent viscosities and
for Weissenberg numbers of order unity or larger. 
Similar final flow states are reached in cases whether the fluid is pre-stressed and close to yielding initially, or if the polymer stress is initially zero.
These states do not depend on the manner in which
small-amplitude perturbations are seeded into the initial conditions.
Thus, the generation of complex patterns around the yield surfaces is not reliant on a specific initial configuration or on having an
underlying base flow that might form stress jumps.
This result is perhaps surprising given that
  the plug persists throughout the evolution, retaining a memory
  of its stress history.

Persistent, complex nonlinear flow states have been reported
previously in other elasto-viscoplastic flows
\citep{devita18,parvar24,mousavi24,abdelgawad24,dzanic24,from25,abdullah26}.
These papers do not provide insight into the mechanism
driving the nonlinear flow states, other than
arguing that they may arise from a plastic modification 
of a known viscoelastic flow instability. This contrasts with our demonstration of a route to complexity via a new type of instability, driven by an interplay between elasticity and plasticity.
Moreover, the flow configurations,
observed dynamics, and geometry of the plugged regions in these studies 
all appear very different to ours, and in some cases \citep{from25,abdullah26} a much higher Weissenberg is required to generate complex dynamics. 

In order to ease simulations,
we have added stress diffusion to
Saramito's model. 
The addition of a small level of stress diffusion is a commonly used
device in viscoelastic flow simulation
\citep[e.g., ][]{dubief23,lellep24,lewy26},
where in some cases it has been justified physically for polymer solutions on the grounds of
kinetic theory. Whether a simple diffusion term is also appropriate
for elasto-viscoplastic materials is unclear, but its
inclusion amounts to a simple modification of the model in the sense
of a convenient regularization. Indeed,
we have demonstrated how stress diffusion smooths the
stresses of uni-directional base flows and suppresses linear instability at
the shortest wavelengths, two features that might otherwise lead one to
question the physical validity of Saramito's constitutive model.
However, stress diffusion can have a pervasive effect
over long times, raising the plug stress of the
base states towards the yield stress and reducing the complexity
of the nonlinear states that result from linear instability.

\cite{evpoiy} previously discussed the mechanism behind the
linear instability in the limit of high Weissenberg number,
pinpointing the streamwise straining and cross-stream shearing
of the base-state normal stress $\Txx$ over regions with negligible
background shear. For finite Weissenberg number and short
wavelength, however, the story becomes
obscured by the localisation of perturbations to the yield surfaces.
Nevertheless, the character of the instablity distinguishes it
from other, known viscoelastic or elastic flow instabilities.
The base state need not develop normal stress discontinuities
and polymer stress diffusion is not essential,
unlike in the viscoelastic flow instabilities reported by
\citet{hinch92} or \cite{beneitez23,beneitez25}. Neither the Weissenberg number nor the solvent viscosity ratio must take extreme values, counter to
the centre-mode instability in Oldroyd-B channel flow \citep{khalid21}.
In the inertia-less instability of viscous fluid flow
over an elastic solid \citep{kumaran2021stability},
linear instability has finite wavenumber
and the interface is material, not a contour of constant stress. 
The closest connection may be with viscoelastic instabilities driven by
strong shear-thinning, which also have a short-wave form
and are associated with regions in the base flow with low shear rate
\citep{wilson99,castillo18}. However,
in those viscoelastic flows, there is no true yield
surface or plug, both of which are key contributors to
our elasto-viscoplastic instability and the complex nonlinear states
that are generated.

\begin{bmhead}[Acknowledgements]
We thank Theo Lewy, Hugo Fran\c{c}a,
Rich Kerswell, Maziyar Jalaal and Mark Martinez
for discussions, suggestions
and assistance in implementing the Dedalus code.
JDS was supported by a Leverhulme Trust Study Abroad Studentship.
NJB acknowledges NSERC Discovery Grant RGPIN-2022-03705.
\end{bmhead}

\begin{bmhead}[Declaration of Interests]
The authors report no conflict of interests.
\end{bmhead}

\appendix

\section{Numerical details}\label{app:num}

We solve \eqref{conteq}-\eqref{bcs} pseudo-spectrally in Dedalus \citep{burns2020dedalus}, typically using $N_x=256$ Fourier modes and $N_y=384$ Chebyshev modes
to discretise in space. We employ dealising with the standard $3/2$ rule, and use a uniform time step, typically $dt=10^{-3}$. In practice, to implement these equations in Dedalus we replace $Y$ in \eqref{Ydef}
with the smoothed version, 
\begin{equation}
	\tilde{Y} = \frac{1}{2}\left[1 + \tanh\left(\frac{\cT - \Bi}{\epsilon}\right)\right]\left(1-\frac{\Bi}{\cT}\right),
\end{equation}
where $\epsilon\ll1$. We use $\epsilon=10^{-4}$ in all simulations presented,
which is small enough to ensure that $\tilde{Y}$ closely approximates $Y$. 

\begin{figure}[t!]
\centering
\includegraphics[width=0.8\textwidth]{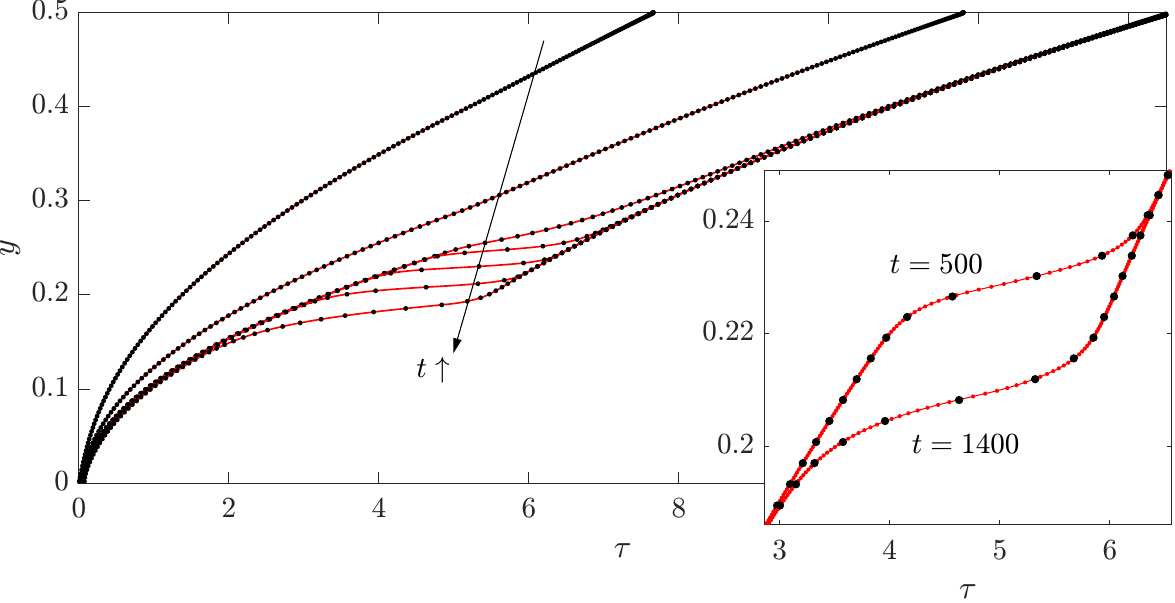}
\caption{Snapshots of $\tau$ at $t=\{10,20,70,150,500,1400,4600\}$ for a
  solution to the uni-directional base-state equations \eqref{315}-\eqref{321}
  solved using ode15s in Matlab (solid red, on a uniform
  spatial grid with 2000 points) and a quasi-1D Dedalus simulation
  (black dots, plotted at the $N_y=384$ Chebyshev nodes)
  with $(N_x,N_y)=(2,384)$. In both cases,
  $(\Bi,\Wi,\beta,\kappa)=(5,10,0.5,10^{-6})$ and there is zero initial
  polymer stress. 
  The inset shows a magnification around the yield surface for the
  two times indicated, with the small red and larger black
  points indicating both numerical grids.
  }
\label{fig:bcheck}
\end{figure}

A first simple numerical test is to compare an effectively 1D Dedalus
simulation, with $N_x=2$ and $\epsilon=10^{-4}$, to
the corresponding solution of the uni-directional base state equations
\eqref{315}-\eqref{321} using MATLAB's ode15s with centred
differences on a much finer, spatially uniform grid.
Figure \ref{fig:bcheck} presents such a comparison,
in which the Dedalus simulation employs $N_y=384$ Chebyshev modes
and the ode15s solution uses 2000 gridpoints.
A small level of polymer diffusion,
$\kappa=10^{-6}$, ensures that stress discontinuities do not develop, and
the spatial resolution of
the Dedalus solution is sufficient to
capture the relatively steep gradient in $\tau$ around
the yield surface (even with relatively few Chebyshev nodes),
and the slow drift of its position with time.

\begin{figure}[t!]
\centering
\includegraphics[width=\textwidth]{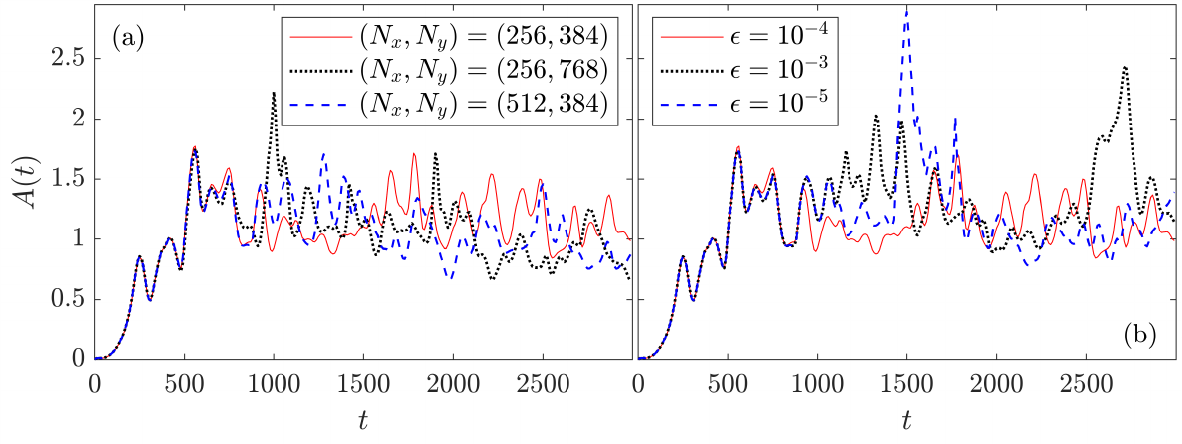}
\caption{(a) Time series of $A(t)$ 
  for simulations with the resolutions indicated,
  for the parameter settings of Fig.~\ref{fig:sim}(a).
  (b) Similar time series, but for simulations with
  the values of $\epsilon$ indicated.
}
\label{fig:grid}
\end{figure}

Fig.~\ref{fig:grid}(a) demonstrates that, for the example simulation presented
in Fig.~\ref{fig:sim}(a), the effect of varying spatial resolution
has no impact, other than triggering
a variation in the solution after what we interpret to be the
Lyapunov time of the underlying chaotic dynamics.
The saturated amplitudes in \eqref{Aend} are measured
to be
\be
\Aend = \{1.064\pm0.355,1.008\pm0.363,1.006\pm0.346\}
,
\ee
where the errorbars denote the standard deviation of each time series
of $A(t)$. Similarly, taking $\epsilon = 10^{-3}$ or $10^{-5}$
instead of $\epsilon=10^{-4}$ has no systematic observable impact
on the dynamics (see Fig.~\ref{fig:grid}(b)),
with the saturated amplitudes measured as
\be
\Aend = \{1.207\pm0.446,1.120\pm0.443\}
,
\ee
respectively.
In all these cases, the values of $\Aend$ differ from one another by
amounts that are significantly less than the natural variability
in the system, as highlighted by the standard deviations
(an intermittent burst of activity occurring in the simulations with
$\epsilon = 10^{-3}$ and $10^{-5}$ enhances the $\Aend$ values
for those cases, a feature that we attribute to relatively
infrequent chaotic excursions over extremities of the strange attractor).

\begin{figure}[t!]
\centering
\includegraphics[width=0.99\textwidth]{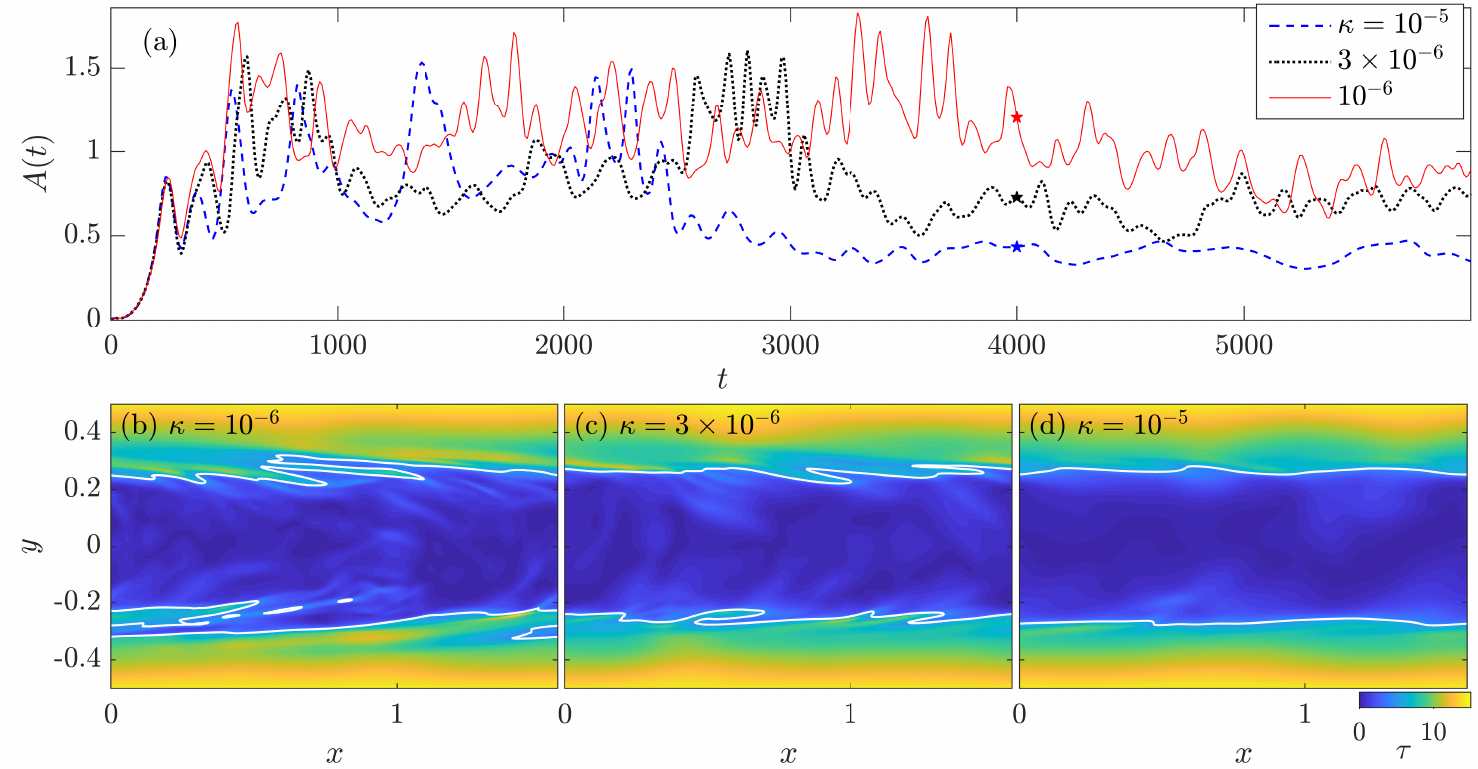}
\caption{
  (a) Time series of $A(t)$ from simulations
  with the diffusivities indicated, but otherwise the same
  parameter values as in figure \ref{fig:sim}(a). (b,c,d) Snapshots
  of $\tau$ at $t=4000$ (the stars in (a)). 
}
\label{fig:diff}
\end{figure}

Our numerical simulations incorporate polymer stress diffusion to ease the
computations and prevent the build of power at the shortest spatial scales.
Proceeding to longer times, however, is problematic
as the simulations become sensitive to the degree of stress diffusion:
Fig.~\ref{fig:diff}(a)
shows time series of $A(t)$ from simulations with different values
of $\kappa\in[10^{-5},10^{-6}]$, each run to a final time of $t=6000$.
For these solutions, stress diffusion has a minor impact 
on the early-time growth rate and $A(t)$ saturates at similar levels.
However, at later times
the level of complexity of the nonlinear state becomes
sensitive to $\kappa$, with the level of $A(t)$ falling to lower values at
earlier times when $\kappa$ is increased. Simultaneously, and as shown by
the snapshots of $\tau$ at $t=4000$, the complex nonlinear state possesses
weaker spatial variation and less deformation of the yield surfaces. 

\bibliographystyle{jfm}
\bibliography{../../../../vpthread/jfm}

\end{document}